\documentclass[rnote]{aa}

\usepackage{graphicx}
\usepackage{txfonts}
\usepackage{natbib}        
\bibpunct{(}{)}{;}{a}{}{,} 

\newcommand{\Ra}{\ensuremath{R_\mathrm{a}}}
\newcommand{\Rb}{\ensuremath{R_\mathrm{b}}}

\newcommand{\Msun}{\ensuremath{\mathrm{M_{\sun}}}}

\newcommand{\Rsun}{\ensuremath{\mathrm{R_{\sun}}}}
\newcommand{\mas}{\ensuremath{\mathrm{\, mas}}}

\newcommand{\haro}{Haro~1-14C}

\begin{document}
\title{Refined masses and distance of the young binary Haro 1-14 C\thanks{Based on observations collected under program 091.C-0093(A) with the PIONIER/VLTI instrument at the European Southern Observatory, Paranal, Chile.}}
\titlerunning{Haro 1-14 C with PIONIER}
\author{
  J.-B.~Le~Bouquin\inst{1}
  \and J.-L.~Monin\inst{1}
  \and J.-B.~Berger\inst{2}
  \and L.~Prato\inst{3}
  \and M.~Benisty\inst{1}
  \and G.~Schaefer\inst{4}
}

\institute{
  UJF-Grenoble 1 / CNRS-INSU, Institut de Plan{\'e}tologie et d'Astrophysique de Grenoble (IPAG) UMR 5274, Grenoble, France
\and
  European Southern Observatory, Garching by M{\"u}nchen
\and
  Lowell Observatory, 1400 West Mars Hill Road, Flagstaff, AZ 86001, USA
\and 
The CHARA Array of Georgia State University, Mount Wilson Observatory, Mount Wilson, CA 91023, USA
}

\offprints{J.B.~Le~Bouquin\\
  \email{jean-baptiste.lebouquin@obs.ujf-grenoble.fr}}
\date{Received 28/10/2013; Accepted 13/12/2013}
\abstract
{}
{We aim to refine the dynamical masses of the individual component of the low-mass pre-main sequence binary \haro{}.}
{We combine the data of the preliminary orbit presented previously with new interferometric observations obtained with the four 8m telescopes of the Very Large Telescope Interferometer.}
{The derived masses are $M_a=0.905\pm0.043\,\Msun$ and $M_b=0.308\pm0.011\,\Msun$ for the primary and secondary components, respectively. This is about five times better than the uncertainties of the preliminary orbit. Moreover, the possibility of larger masses is now securely discarded. The new dynamical distance, $d=96\pm\,9\,$pc, is smaller than the distance to the Ophiuchus core with a significance of $2.6\,\sigma$. Fitting the spectral energy distribution yields apparent diameters of $\phi_a=0.13\pm0.01\mas$ and $\phi_b=0.10\pm0.01\mas$ (corresponding to $\Ra=1.50\,\Rsun$ and $\Rb=1.13\,\Rsun$) and a visual extinction of $A_v\approx1.75$. Although the revised orbit has a nearly edge-on geometry, the system is unlikely to be a long-period eclipsing binary.}
{The secondary in Haro~1-14C is one of the few low-mass, pre-main sequence stars with an accurately determined dynamical mass and distance.}

\maketitle

\section{Introduction}

\haro{} (\object{HBC 644}) is one of the few low-mass, pre-main sequence binaries whose orbit can be observed both with radial velocity and resolved astrometric measurements. This provides a way of measuring the dynamical masses and distance and of testing the theoretical evolutionary models. Consequently, \haro{} attracted the attention of observers over the past decade. \cite{Reipurth:2002} first identified the system as a single-lined spectroscopic binary. \citet{Simon:2004} detected the radial velocity of the secondary and determined the full double-lined spectroscopic orbit. Finally, \citet{Schaefer:2008} spatially resolved the system for the first time using optical long-baseline interferometry and derived a preliminary three-dimensional orbit.

The apparent magnitudes of \haro{} in the near-infrared are $8.0\pm0.05\,$mag and $7.78\pm0.03\,$mag in the H and K bands, respectively (2MASS Catalog). This is too faint for the existing suit of instruments installed at the focus of imaging interferometers, which operate with 1-meter class telescopes (such as CHARA or VLTI-AT). Observations from \citet{Schaefer:2008} were obtained with the two 10-meter telescopes of the Keck Interferometer (Keck-I). However, the single Keck-I baseline provides poor sampling in (u,v) coverage. This is especially true for \haro{}, given its declination. Consequently, the preliminary three-dimensional orbit suffers from a long tail of possible solutions in the probability distribution toward smaller inclinations and toward larger masses for the components: $M_a=0.96^{+0.27}_{-0.08}\Msun$, $M_b=0.33^{+0.09}_{-0.02}\Msun$ and $d=111^{+19}_{-18}\,$pc.

\section{Observations}
\label{sec:observation}

\begin{figure*}
  \centering
  \includegraphics[width=0.9\textwidth]{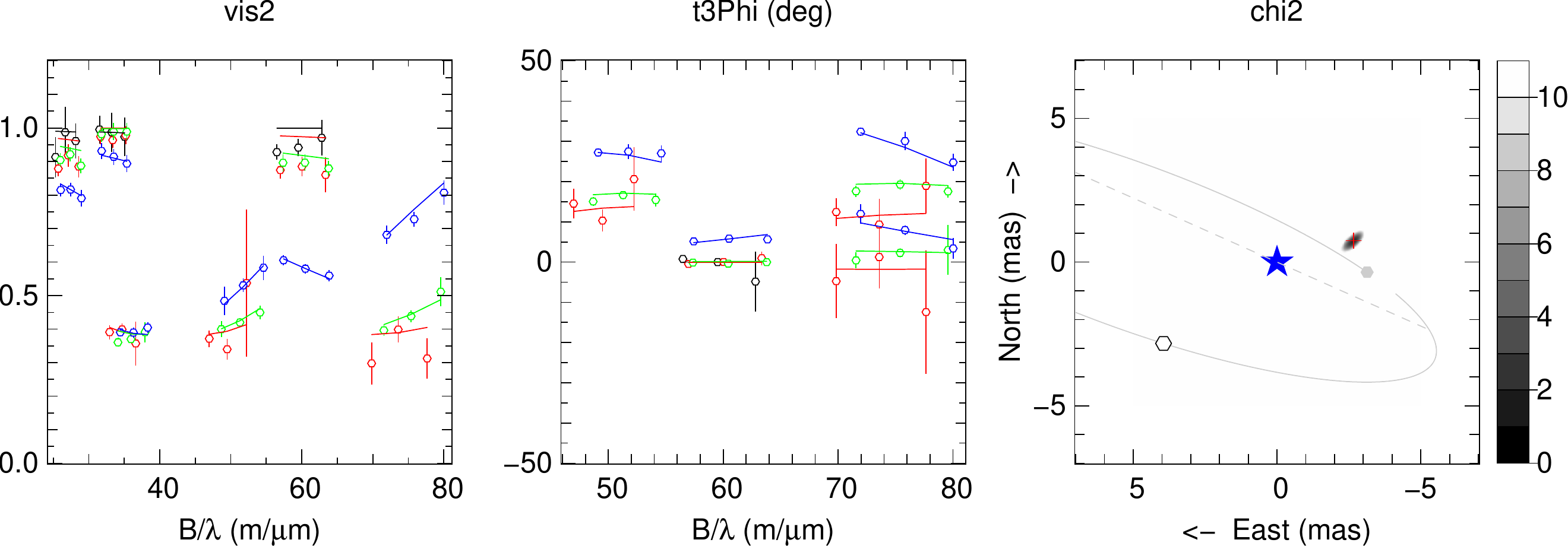}
  \caption{Left and middle panels: PIONIER square visibilities and closure phase (dots) plotted together with the best-fit binary model (solid lines). The colors represent the four different pointings detailed in Table~\ref{tab:obs}. The right-most panel shows the reduced $\chi^2$ of the model with respect to the position of the secondary. The red cross represents the astrometry of the best-fit model, while the open circle represents the prediction from the published orbit at the time of observation. The gray line indicates the trace of the orbit, the gray circle its periastron passage, and the gray dashed line shows the line of nodes.}
  \label{fig:fit_interfdata}
\end{figure*}

Supplementary interferometric data were obtained with the PIONIER combiner \citep[][]{Le-Bouquin:2011} at the Very Large Telescope Interferometer \citep[VLTI,][]{Haguenauer:2010}. The use of the four 8-meter unit telescopes (instead of the four 1.8-meter auxiliary telescopes) was necessary to observe \haro{} (H$=$8.0 mag). The visible magnitude (V$=$12.3 mag) was enough for the MACAO adaptive optics system to provide an adequate correction during the observations. Data were dispersed over three spectral channels across the H band. We obtained four calibrated points during the night 2013-06-19 (ESO convention). Telescope UT4 was missing during the first point because of technical problems. To calibrate the fringe visibilities and the closure phase, we interleaved observations of unresolved single stars between those of \haro{}. These reference stars were found with the \texttt{SearchCal}\footnote{http://www.jmmc.fr/searchcal\_page} software in its faint mode \citep{Bonneau:2011a}. Table~\ref{tab:obs} summarizes the log of observations. Data were reduced and calibrated with the \texttt{pndrs} package \citep[][]{Le-Bouquin:2011}.

\begin{table}[h]
\caption{Log of observations with the Uniform Disk Diameter (UDD) of the reference stars.}
\centering
\begin{tabular*}{0.95\columnwidth}{llll}
\hline\hline\noalign{\smallskip}
MJD  & Target  & Tel. used & UDD \\
\noalign{\smallskip}\hline\noalign{\smallskip}
56463.0268 & HD147284    & 1-2-3-4 & $0.15\mas$ \\
56463.0438 & Haro 1-14C  & 1-2-3   &      \\
56463.0560 & HD147935    & 1-2-3-4 & $0.13\mas$ \\
56463.0683 & Haro 1-14C  & 1-2-3-4 &      \\
56463.0777 & HD147284    & 1-2-3-4 & $0.15\mas$ \\
56463.0874 & Haro 1-14C  & 1-2-3-4 &      \\
56463.0971 & HD147137    & 1-2-3-4 & $0.11\mas$ \\
56463.1601 & HD147284    & 1-2-3-4 & $0.15\mas$ \\
56463.1788 & Haro 1-14C  & 1-2-3-4 &      \\
56463.1894 & HD147935    & 1-2-3-4 & $0.13\mas$ \\
\noalign{\smallskip}\hline
\end{tabular*}
\label{tab:obs}
\end{table}

The calibrated square visibilities and closure phases are shown in Fig.~\ref{fig:fit_interfdata} together with the best-fit binary model. The expected diameters of the individual components, \mbox{$<0.15\mas$}, are unresolved by the longest VLTI baselines and were set to zero in the fitting procedure. Thanks to the simultaneous combination of four telescopes, the binary was detected without ambiguity in the relative position of the two components. We performed a bootstrap analysis by randomly selecting subsets of the observations and adding noise to these datasets. The uncertainty is given by the dispersion of the best-fit parameters when fitting these datasets. The best-fit position of the faint component with respect to the bright component is ($\delta{}\mathrm{East}=-2.66\mas$, $\delta{}\mathrm{North}=+0.74\mas$). The $3\sigma{}$ error ellipse is ($0.53\mas$, $0.22\mas$) with a position angle for the major axis of $130\deg$, measured east from north. The average Modified Julian Day of observation is $56463.104$.

\section{Combined orbit}
\label{sec:orbit}

\begin{figure*}
  \centering
  \includegraphics[width=0.94\textwidth]{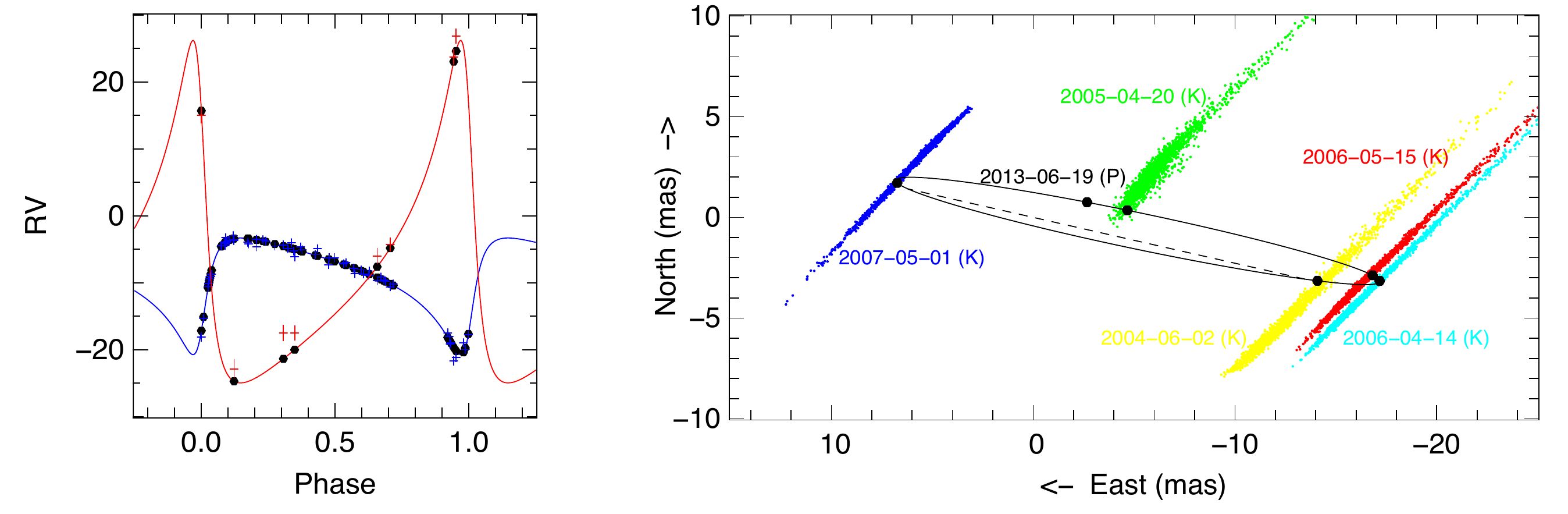}
  \caption{Left: the best-fit orbital solution compared with radial velocities of the primary (blue) and of the secondary (red). Right: the best-fit orbital solution compared with the pseudo-astrometric uncertainty regions from Keck-I (colors) and VLTI (black, smaller than the symbol size). In both plots the black circles are the predictions from the best-fit orbital solution.}
  \label{fig:orbit}
\end{figure*}

\begin{figure*}
  \centering
  \includegraphics[width=0.9\textwidth]{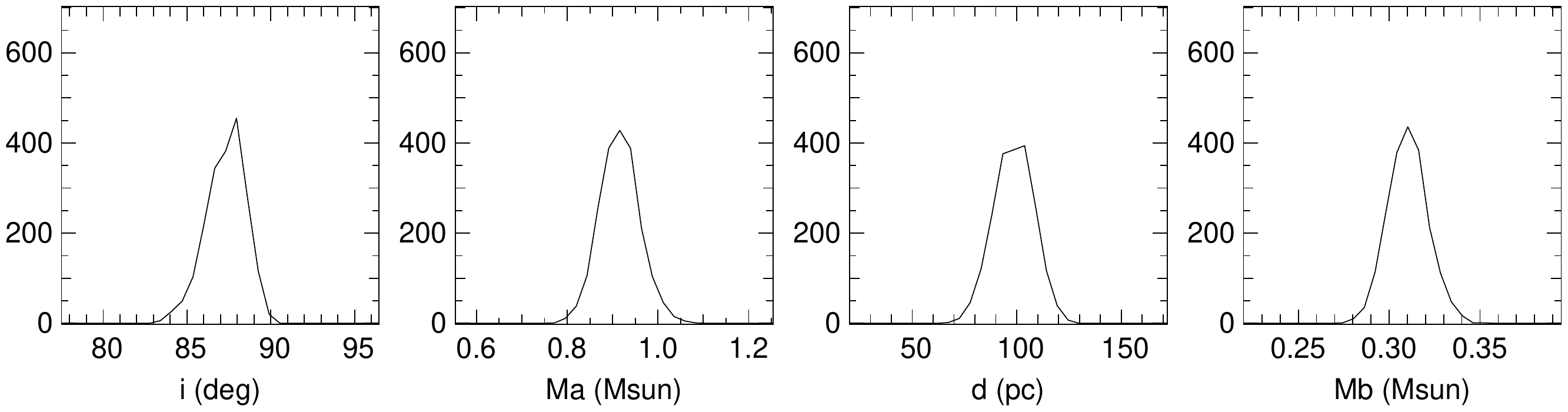}
  \caption{Probability distributions for the most relevant parameters (inclination, distance, and masses of the components).}
  \label{fig:proba}
\end{figure*}

At first consideration, the orbit published by \citet{Schaefer:2008} is largely inconsistent with the PIONIER astrometry (reduced \mbox{$\chi^2_r=122$}). The observed astrometry is in the vicinity of the periastron, but the date of observation is far from the expected periastron date (MJD=56616.1), which is well constrained by the radial velocities of the primary.

However, the Keck-I observations are insensitive to the absolute orientation because of the lack of closure-phase measurements: the published orbit can be flipped by $\Omega'=\Omega+180\deg$. Doing so yields a better agreement and a discrepancy now oriented along the large uncertainties of the preliminary orbit. Starting from this flipped orbit, we ran a combined fit on all available data: radial velocities, Keck-I square visibilities, and PIONIER closure phases and square visibilities. All orbital elements were let free to vary, but this new fit mainly updates the inclination. The best-fit orbital elements are summarized in Table~\ref{tab:bestorbit} and compared with the preliminary solution from \citet{Schaefer:2008}.

\begin{table}[t]
\caption{Best-fit orbital elements and related physical parameters}
\centering
\begin{tabular*}{0.95\columnwidth}{llll}
\hline\hline\noalign{\smallskip}
Param.  & Schaefer, 2008  & This work & Unit\\
\noalign{\smallskip}\hline\noalign{\smallskip}
$P$        & 592.11    $\pm$  0.12    & 592.19   $\pm$ 0.09    & days   \\      
$T$        & 53655.51  $\pm$  0.86    & 53656.22 $\pm$ 0.72    & MJD    \\      
$e$        & 0.6211    $\pm$  0.0070  & 0.617     $\pm$ 0.006    &       \\     
$\omega$   & 232.4     $\pm$  1.2     & 233.56   $\pm$ 0.56 & $\deg$ \\      
$a$        & $13.0^{+2.5}_{-0.7}$       & 15.4 $\pm$ 1.5 & $\mas$  \\     
$i$        & $75.9^{+9.3}_{-25.3}$      & 86.9 $\pm$ 1.2 &  $\deg$  \\     
$\Omega\;^\mathit{[a]}$   & $246^{+18}_{-12}$          & 263.5 $\pm$ 7.1     &$\deg$       \\      
$K_a$      & 8.73      $\pm$  0.13    & 8.71 $\pm$ 0.11  & km/s          \\      
$K_b$      & 25.53     $\pm$  0.56    & 25.59 $\pm$ 0.49 & km/s          \\      
$\gamma\;^\mathit{[b]}$   & -8.788    $\pm$  0.076   & -8.81 $\pm$ 0.07 & km/s        \\      
$f_H$  &                          & 0.239 $\pm$ 0.004 & H-band    \\     
$f_K$  & $0.268^{+0.030}_{-0.048}$  & 0.22  $\pm$ 0.03  & K-band    \\     
\noalign{\smallskip}\hline\noalign{\smallskip}
$M_a$      & $0.975^{+0.957}_{-0.073}$   & 0.905 $\pm$ 0.043 & \Msun     \\      
$M_b$      & $0.333^{+0.327}_{-0.025}$   & 0.308 $\pm$ 0.011 & \Msun     \\  
$d$        & $116.^{+38}_{-21}$         & 95.55 $\pm$ 9 & pc       \\   
\noalign{\smallskip}\hline\noalign{\smallskip}
$\chi^2_r$ & 1.23 & 1.33                           \\
\noalign{\smallskip}\hline
\end{tabular*}
\tablefoot{$\;^\mathit{[a]}$: The $\Omega$ value from Schaefer is the one of the swapped orbit ($+180\deg$), not the published orbit. See section~\ref{sec:orbit}. $\;^\mathit{[b]}$: The systemic velocity is defined in the heliocentric reference frame.}
\label{tab:bestorbit}
\end{table}

We performed a simple bootstrap analysis by adding noise to the data (radial velocities, visibilities, and closure phases) according to the uncertainties, and fitting these noisy datasets. For illustration purposes, we computed the binary separations at the times of the interferometer observations to build the pseudo-astrometric uncertainty regions shown in Fig.~\ref{fig:orbit}. We discarded the VLTI observation when building the Keck pseudo-astrometry, and vice versa. The best-fit orbital solution and its uncertainties were computed with the entire dataset. The probability distribution for the most relevant parameters (inclination, distance, and masses) are shown in Fig.~\ref{fig:proba}. The preliminary orbit suffers from a long tail in the probability distribution toward smaller inclinations as a direct consequence of the almost linear arrangement of all observed (u,v) points. This is not true anymore with the addition of the PIONIER observations: the probability distributions can be approximated by normal distributions. The uncertainties presented Table~\ref{tab:bestorbit} are the standard deviation of the probability distribution of the corresponding parameter in the bootstrap analysis. The relative uncertainties on the stellar masses are now $\pm5\%$ and $\pm3.5\%$ for the primary and the secondary component, respectively. This is about five times better than the uncertainties of the preliminary orbit. Moreover, the tail toward larger masses in the probability distributions is now securely discarded.

The revised orbital solution yields a dynamical distance of $d=96\pm\,9\,$pc. Interestingly, this value is lower than the distance to the Ophiuchus core \citep[$119\pm4\,$pc,][]{Loinard:2008} with a significance of $2.6\,\sigma$.

\section{Discussion}
\label{sec:discussion}

\begin{figure}
  \centering
  \includegraphics[width=0.38\textwidth]{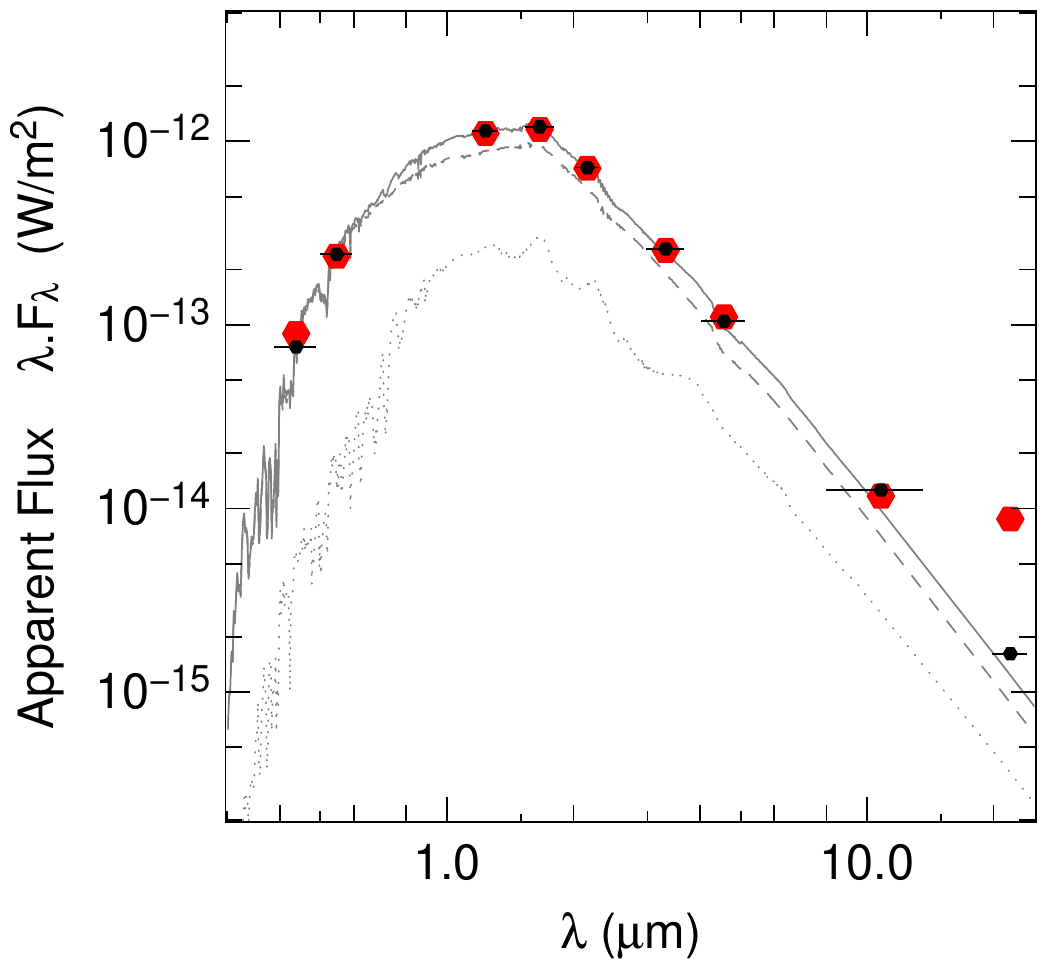}
  \caption{Best-fit synthetic fluxes (horizontal bars) to the spectral-energy distribution from NOMAD, 2MASS, and WISE (red circles). The corresponding synthetic spectra for the two individual components and their sum are overlaid (dashed, dotted, and solid gray lines).}
  \label{fig:sed}
\end{figure}

The secondary in \haro{} is one of the few low-mass pre-main sequence stars with an accurately determined of its dynamical mass and distance. The remaining main limiting factor in making comparisons with the evolutionary tracks is the estimation of the effective temperature. Given the binary separation, its period and total brightness, GAIA should be able to accurately measure the flux ratio in the visible. Combined with the flux ratio in the near-infrared, this provides key information for separating the spectral energy distribution of the secondary, and possibly for estimating its effective temperature with respect to that of the primary. It will also be possible to compare the observations with models using the V$-$K colors of each component in place of effective temperatures. This approach is potentially more precise \citep{2013ApJ...773...40T}.

Another possible solution lies in a direct measurement of the component diameters. However, they are out of reach of current optical interferometers with hectometric baseline lengths. Given the high inclination of the revised orbit, one might wonder whether \haro{} might be a long period eclipsing binary. To test this idea, we fitted the spectral energy distribution with the sum of two atmospheric models (see Fig.~\ref{fig:sed}). We used effective temperatures of $4400\pm200$\,K for the primary and $3500\pm200$\,K for the secondary \citep[based on the spectral types from][]{Schaefer:2008}. The interferometric flux ratio in the near-infrared corresponds to a ratio of $0.75\pm0.1$ for the linear radii. We then adjusted the photometry by varying the apparent diameters and the visual extinction $A_V$, assuming that the latter is the same for the two components. The W4 photometric point at $22\,\mu$m was discarded from the fit because it is a clear outlier in the Rayleigh-Jeans regime. The best match is obtained for $A_V=1.8\pm0.7$ and for $\phi_a=0.131\pm0.015\mas$ and $\phi_b=0.098\pm0.011\mas$ using \texttt{ATLAS9} models \citep{Castelli:2004}. Differences using \texttt{NextGen} models \citep{Allard:1997} are negligible. Unfortunately, \haro{} cannot be an eclipsing binary because the closest approach between the two components in the apparent orbit is $0.32\mas$. These diameters correspond to $R_a=1.50\,\Rsun$ and $R_b=1.13\,\Rsun$ in linear radius at the distance of the object.

Finally, it would be very interesting to confirm the distance discrepancy between \haro{} and the core of the Ophiuchus cloud. The dynamical distance is the physical parameter that is the least constrained by the existing dataset. This is because the distance linearly depends on the size of the apparent orbit ($a_{app}$), which is still degenerate at the 10\% level with the orientation on sky ($\Omega$). A few additional, high-quality astrometric observations at various epochs are needed to lift this degeneracy.

\listofobjects{}

\begin{acknowledgements} 
PIONIER is funded by the Universit\'e Joseph Fourier (UJF), the Institut de Plan\'etologie et d'Astrophysique de Grenoble (IPAG), the Agence Nationale pour la Recherche (ANR-06-BLAN-0421 and ANR-10-BLAN-0505), and the Institut National des Science de l'Univers (INSU PNP and PNPS). The integrated optics beam combiner is the result of a collaboration between IPAG and CEA-LETI based on CNES R\&T funding. The authors warmly thank everyone involved in the VLTI project. This work is based on observations made with the ESO telescopes. It made use of the Smithsonian/NASA Astrophysics Data System (ADS) and of the Centre de Donnees astronomiques de Strasbourg (CDS). All calculations and graphics were performed with the freeware \texttt{Yorick}. The work of L.P. was supported by NSF grant AST-1009136.
\end{acknowledgements}


\end{document}